# Travel Guides for Creative Tourists, Powered by Geotagged Social Media

**Dan Tasse, Jason I. Hong**
Carnegie Mellon University
Pittsburgh, PA
jasonh@cs.cmu.edu

**ABSTRACT**
Many modern tourists want to know about everyday life and spend time like a local in a new city. Current tools and guides typically provide them with lists of sights to see, which do not meet their needs. Manually building new tools for them would not scale. However, public geotagged social media data, like tweets and photos, have the potential to fill this gap, showing users an interesting and unique side of a place. Through three studies surrounding the design and construction of a social-media-powered Neighborhood Guides website, we show recommendations for building such a site. Our findings highlight an important aspect of social media: while it lacks the user base and consistency to directly reflect users' lives, it does reveal the idealized everyday life that so many visitors want to know about.

**ACM Classification Keywords**
H.5.m. Information Interfaces and Presentation (e.g. HCI): Miscellaneous

**INTRODUCTION**
Tourism has seen a demographic shift, with the rise of a group called "creative tourists" [35]. These travelers want to "understand the feel of an area," to "see everyday life," to "live like a local" instead of just "seeing the tourist sights." But in an unfamiliar place, they may wonder where to go. After all, some parts of any city may be boring, dangerous, or otherwise unsuitable. Creative tourists need to be able to understand the neighborhoods of a city to plan out an enjoyable trip.

Travelers have a unique set of information needs, because they are new to a place and do not have time to build up local knowledge from experience. They also have more opportunities of places to stay and visit, thanks to platforms like Airbnb and Couchsurfing. But they do not have the informational tools to understand this wider array of options. Traditional guidebooks from Fodor's, Frommer's, and Lonely Planet focus on central tourist districts and sights to see. Yelp and Foursquare give people information about the bars, restaurants, and shops in an area, but it can be hard for travelers to get an overall sense of how a neighborhood feels from just this information. Cities gather statistics, and are releasing open data more than ever before, but numbers also fail to convey a neighborhood's culture.

At the same time, the rise of social networking means that people have been posting tons of photos, tweets, and checkins with their location attached. (One estimate has at least 500 million tweets sent per day [24], and a recent dataset release from Yahoo included 100 million public photos, of which about half are geotagged [43].) These photos and other posts, all tied to locations, can help travelers understand the cities they are traveling to. Social media posts have advantages over traditional travel guides as well. They are scalable, so it is feasible to build guides cheaply to cover any neighborhood in almost any city, and they are democratic: the resulting view of the neighborhood is controlled not by a publisher or elite critic, but by everyone adding their own experiences.

We aimed to understand two main questions: what do creative tourists mean by "getting a feel for the city", and how can social media help them find it? To answer these, we first built a model of tourist information search based on interviews with 14 travelers and 490 survey responses. Informed by this model, we designed a Neighborhood Guides web app and evaluated it iteratively with 30 participants to understand what it helps them with and what they still need.

The contributions of this paper are therefore not the Neighborhood Guides website itself, but the following insights, which we acquired through the Neighborhood Guides site's iterative design process. We found that creative tourists want safety, convenience, liveliness, aesthetic appeal, and the ability to live like a local. Social media, particularly a diverse and well-organized set of photos selected using their annotations and contents, can help creative tourists find these dimensions they want. Finally, we demonstrate a finding about social media. These geotagged social media posts can offer one lens into the city: they show the idealized city, not the "realistic" one, but this is what these tourists want.

## RELATED WORK

Many researchers have used social media to describe cities. They often run into problems because of some major limitations inherent in social media data. Tourism can best deal with the opportunities and limitations provided by social media, so we will focus on those applications. Even within social media for tourism research, though, we find that some applications are more in line with recent changes in tourism. We describe these changes, particularly the rise of creative tourism, and describe how we aim to address these new creative tourists.

### Describing Cities Using Social Media

Plenty of researchers have attempted to deepen our knowledge of cities with social media in a number of ways.

For example, some have found neighborhood boundaries that reflect social movement based on Foursquare [10, 48] or Twitter [45]. Similar work finds boundaries of informal regions like "downtown" or "red light district" [17, 42], or used social media to show that colloquial distinctions like "9th Street Divide" do not reflect real movement [38]. These are likely too fine-grained for travelers, who would care more about what a region is like than the exact definition of the borders.

Some have addressed the question of "what a region is like" by trying to summarize social media content in the area. Jaffe et al. [20] addressed the problem of summarizing photo content by finding a subset of photos that would accurately summarize a larger photo set. They did this by clustering all of the photos and then ranking the clusters based on five criteria: tag distinguishability, photographer-distinguishability, density, image qualities, and arbitrary relevance factor (such as a search query). Kennedy et al. and Rattenbury et al. [22, 34] further developed the ability to find the "most representative" image, and the most representative image tags, from a set of photos using computer vision features such as SIFT. Crandall et al. [8] did the same: finding the top N "interesting" places in each city and a "canonical" photo from each. Finally, Kafsi et al. further expand this work to understand which tags are locally relevant, which are city-level, and which are country-level [21].

Summarizing textual content, like tweets, is somewhat easier because there is less total information, so one can use a simple method like a word cloud (at least as a supplementary tool) to get a sense of a large corpus of words [30]. More intelligent methods have been used for tweets, for tasks like event detection [25] and location modeling [23]. Importantly for neighborhoods, though, Hao et al. approach high-level neighborhood modeling in another interesting manner, creating Location-Topic Models based on what users write in travelogues [15].

*Challenges in these approaches*

While these approaches all have valid uses and results, they contain some shortcomings. Goodspeed cites three main issues: content poverty, espoused theory vs. theory-in-use, and positivist assumptions [14]. In short, social media data is broad but not deep; we should not conflate the size of data available with the usefulness of that data. As examples of these biases, Schwartz and Hochman [37] describe how social media does not reflect demographics and favors exceptions over daily life, while Hecht and Stephens [16] show that this data is biased towards urban areas and under-represents what happens in rural places.

As a result, we cannot simply use existing work directly to describe a city completely. The biases towards urban and exceptional occurrences, though, may make social media a particularly useful data source for urban tourism, and indeed, many researchers have attempted this. In the next section, we describe some of these attempts.

### Providing Recommendations to Tourists

Using social media to help tourists is not a new idea. Since the early 21st century, researchers have tried to use the abundance of social media data to recommend things for tourists to do. Work in this vein includes recommendations of restaurants [18], shops [41], travel routes [26, 27, 31], attractions and points of interest [2, 13], and destinations [15]. These all use social media and user-generated content such as user locations, so continuing in this vein seems like a logical choice. In addition, sites like Yelp and Foursquare have dozens of user reviews, so aggregating reviews and recommending the most highly-rated spots seems like a natural solution.

However, this approach has three shortcomings. First, people need to know why they are recommended each place. It would be rare for tourists to set out on a trip solely because an algorithm recommended it. Second, they solve problems that are already solved by Yelp and Foursquare: finding a restaurant or a point of interest by consulting one of these guides is easy. Most importantly, though, these projects fail to take into account recent changes in modern tourism. In particular, the rising group of "creative tourists" would not be served by any recommendation service. They want to create their own journey, not have it given to them.

### The Rise of Creative Tourists

These "creative tourists" travel differently than previous tourists. Travel in previous decades had meant traveling to beaches, beautiful natural sites, or resort towns, but in recent years urban tourism is the fastest growing segment of the tourism market [5]. The character of urban tourism is changing as well as the volume: new urban tourists want to "experience and feel a part of everyday life." [29] Furthermore, they seek to have an active hand in co-creating the experiences, rather than passively paying for and absorbing an experience [1]. Lists of sights to see and experiences to buy no longer suffice. In this paper, we will call people who travel in this way "creative tourists," after Richards [35], though others use terms such as "new urban tourists" [12] or "Explorers" [44].

When creative tourists travel to a city, they are often looking for an authentic experience of that city, rather than a manufactured diversion. The search for authenticity in tourism has a long history dating back at least to the 1970s [28], but recent developments have aided this search in new ways, particularly with regard to lodging. Because hotels historically clustered in downtowns, they could not show travelers all the sides of a city, so travelers turned to alternatives. The peer-to-peer lodging rental site Airbnb, for example, has become a popular and "authentic" way for travelers to rent rooms in residential parts of town [40, 47]. Similarly, Couchsurfing allows users to stay with locals for free (often on their spare couch) [47].

Unlike the sun-and-sand tourists of two generations ago or the cultural-site-visiting tourists of last generation [11], these creative tourists want to curate and create their own experience [35]. They want to stay in interesting residential neighborhoods and spend time "wandering about", "taking in the city", and "getting among the people" [1]. To do this, they need guides to areas, not specific venues. Urban tourism depends on the serendipity and spontaneity that results from getting to know neighborhoods, and on the individual's ability to co-create their experience. Current tools help people discover points, not overall pictures of parts of the cities.

**Approaching Social Media for Creative Tourism**
Some recent work has suggested ways that might move beyond the "lists of points" approach into something that might help creative tourists understand a place holistically. Curated City [9] is one such project: in it, users can create a guide to their neighborhood for others. They do so by contributing photos and answers to the prompt "This is my favorite place for ___ in the neighborhood." Airbnb also offers neighborhood guides, created by local writers and photographers [1]. Participatory projects like "Anywhere" [3], in which a participant is guided around a city by an unseen performer, would be another way people could quickly get under the surface of a place.

However, the weakness of these projects is in their scalability. Curated City and Airbnb's neighborhood guides need a great deal of human contribution for each new city they want to add, while participatory projects require people working in real time. As a result, we want to make something that will serve users as well as these guides, but do so automatically. This brings us back to using social media data, because it can address this issue of scale, it is created by people who are in the city instead of outsiders, and it already exists.

**RESEARCH APPROACH**
In this work, we wanted to address two research questions: what do creative tourists want, and how can social media help them find it? We situate our work at the point where a traveler is just starting to plan a trip; after they know what city they are visiting, but before they are considering exact lodging options. Riegelsberger et al. [36] call this Step 0, "Lay of the Land."

We employed a mixed methods approach to answer both questions. For the first, we first used a series of interviews to generate a provisional model for what these tourists are seeking, then used a survey to refine and validate that model. To answer the second question, we built a prototype Neighborhood Guides website and conducted an in-person user study, along with a quantitative study of which photos are most representative of a neighborhood.

**STUDY 1: INTERVIEWS OF CREATIVE TOURISTS**
For this research, we employed a mixed methods approach, gaining qualitative insights from interactive interviews, which we adjusted and confirmed with quantitative data from surveys (described later as Study 2). We began with interviews of 14 participants, each about a recent trip they took.

---
[1] http://www.airbnb.com/locations

We recruited 7 participants in Pittsburgh who all recently traveled by posting our study on Reddit, Craigslist, and Facebook. We asked for people who preferred to see the "everyday" side of a place instead of just "seeing the sights"; as there is no concrete definition of "creative tourists", we did this to recruit them as well as possible. We asked them to describe their search process and their experience finding a neighborhood to stay. We also recruited seven more recent travelers in San Francisco, bringing the total to 14 participants. Our university's Institutional Review Board approved this study.

We will refer to the first seven interviewees as A1-A7, and the next seven as B1-B7. Interviews were conducted in cafes or other public places near them for convenience and to get them thinking about their neighborhoods. B5 and B6, a dating couple, interviewed together; all the rest were done separately. Our participants are mostly in their 20s and 30s, which are also the age groups most likely to try a home-sharing site like Airbnb or Couchsurfing [39].

Because interviews occurred in public places, state law and our institution's IRB dictated that we could not record them due to the possibility of incidentally recording conversations from others who had not consented. Instead, we took detailed notes to capture important points as well as possible. After finishing each batch of interviews, we analyzed the data iteratively, using an open coding approach to allow insights to emerge from the data.

These interviews revealed a lot about this group's travel motivations, what they hope to learn about neighborhoods, and where they decide to stay, as well as a few interesting tensions that arise when they make those decisions. We built a six-part model of tourist information search, which we then refined and verified with our next study; the dimensions were Safety, Diversity, Walkability, Aesthetics, Third Places, and Authenticity. For brevity and clarity, we will discuss the model in more detail after its refinement in Study 2.

**STUDY 2: SURVEY TO REFINE TOURIST NEEDS MODEL**
To refine our model after Study 1, we conducted a survey. We wanted to know if our six-dimensional model was complete, if we missed any dimensions, and if all six were necessary.

The survey asked participants which of our six dimensions was the most important thing when they are finding a place to stay (with an option for "other"), then asked them more nuanced questions about the relative importance of each one. These questions were taken from key points people brought up during the interviews; for each of our six dimensions, we created 2-4 questions based on that dimension. The survey questions are shown in Table 1.

We recruited participants in two batches. In the first batch, we recruited on Facebook, Twitter, Reddit, Craigslist, and Slack, and through a participant pool at our university. The survey took about 10 minutes, and participants were entered into a drawing for one of five $50 Amazon.com gift cards. We

| Dimension | | Question text |
|---|---|---|
| | 1 | Which of the following is the most important to you when finding a neighborhood to stay in when you travel? (Safety; Diversity of people there; Walkability; Aesthetic appeal; Cafes, bars, and social spaces; Authenticity; Other (enter your answer)) |
| Safety | 2 | How concerned are you with safety when you travel? |
| | 3 | How influential is an area's crime rate in deciding where you will stay? |
| Diversity | 4 | When you travel, how desirable is it for people who live in the area you're staying in to be diverse? |
| | 5 | How often do you go to places where lots of different people interact? |
| | 6 | Would you rather stay in an area that is "up and coming" or an area that is "established"? |
| Walkability | 7 | How important is it to be able to get around by walking when you travel? |
| | 8 | When you travel, how desirable is it to be in an urban place with lots of activity? |
| | 9 | How important is it to be able to get around with public transit when you travel? |
| | 10 | How often do you have a car (whether you drive it to your destination or rent it there) when you travel? |
| Aesthetics | 11 | When you travel, how important is it that the neighborhood you stay in looks nice? |
| | 12 | How influential is the "look" of a neighborhood in choosing where you want to stay? |
| 3rd places | 13 | How important is it that the neighborhood you stay in has great bars, cafes, or other social spaces? |
| | 14 | When you travel, how often do you speculate what life would be like if you lived there? |
| | 15 | When you travel, how often do you try to do what the locals do? |
| Authenticity | 16 | When you travel, how desirable is it to stay in an area that caters to tourists? |
| | 17 | How important is it to you to find "off the beaten path" places when you travel? |
| | 18 | How important is it that the neighborhood you stay in is also a functional neighborhood for people who live there? |

**Table 1. Survey questions for Study 2.** All questions after question 1 were presented in random order and had 5-point Likert scale responses. "How desirable" questions (4, 8, 16) had responses from "Very undesirable" to "Very desirable." "How Important," "How Influential," and "How Concerned" questions (2, 3, 7, 9, 11, 12, 13, 17, 18) had "Not at all/Slightly/Somewhat/Very/Extremely" responses. "How often" questions (5, 10, 14, 15) had "Never", "Almost never", "Occasionally", "Almost every time", "Every time" responses. Question 6 had "Much prefer up and coming", "Somewhat prefer up and coming", "Neutral", "Somewhat prefer established", "Much prefer established" responses.

received 98 responses. This survey was approved by our Institutional Review Board. For the second batch, our colleagues at Airbnb sent the survey to some of their users and received 392 responses. We chose to recruit from Airbnb because it is a popular site among creative tourists [40, 47].

**STUDIES 1 AND 2 RESULTS**
A number of conditions may cause travelers to do very little research before choosing where to stay. If someone already has a place to stay, they will likely take that. B2 described this as a "bird in the hand" situation, and said it occurred a lot when Couchsurfing: finding a local who's willing to host him for free can be difficult, so he will usually accept, regardless of circumstances.

If a traveler has social or other constraints, such as friends or family to visit or an event to attend, they usually consider tourism secondary and stay somewhere nice near that constraint. B5 and B6 described going to the X Games, an extreme sports event, in Aspen, Colorado: they spent most of their time watching events, so they simply wanted to stay near the games. Similarly, B4 described visiting Scottsdale, Arizona, on personal business, which led to him staying in the Fashion Square district. He found it rather unpleasant, and had trouble getting around, but he needed to be near there.

Finally, budget constraints would often short-circuit the lodging search. B5 and B6 described another trip, when they went to Seattle but wanted to pick the cheapest lodging possible. This ended up being the Green Tortoise Hostel downtown, and since they had stayed in another Green Tortoise elsewhere, they decided it would work. B3 also described a road trip where he simply looked up a place to stay while on the road each day, only wanting something simple, clean, and cheap.

**The five dimensions of creative tourist information search**
Most of the participants in Study 1 described at least some trips where they did not use any of these heuristics, and instead wanted to satisfy six different dimensions: Safety, Diversity, Aesthetic Appeal, Walkability, Third Places, and Authenticity.

After Study 2, we adjusted the model and instead found five dimensions. We did this by performing factor analysis on our survey questions. Five eigenvalues of the correlation matrix were greater than 1 in both datasets, so we performed factor analysis with five factors. Loadings are shown in Table 2. We will explain these dimensions in the next subsections.

**Dimension 1: Safety**
Everyone wanted to be safe. Safety was a dimension that came out of our interviews, and our surveys verified its importance. The meaning of safety varied slightly depending on location; usually it included crime, but A1, A7, and B4 all mentioned fear of bedbugs when traveling to New York. People's interpretations of "safety" varied, too, depending on the type of crime and individual thresholds. B4 didn't really care about most crime, except, "I just don't want to get shot at." On the other hand, A2 rerouted a whole itinerary through France after hearing that Marseille was "unsafe."

When asked if safety was always an upside, many participants declined. A2 described spending one night in Churchill Gardens, a posh part of London, but then moving on to somewhat simpler Clerkenwell. Often the safest spaces are also the most expensive, and because they are so expensive, only a homogenous set of wealthy people can live there.

| Question | Factors | | | | |
|---|---|---|---|---|---|
| | 1 | 2 | 3 | 4 | 5 |
| 2 (importance of safety) | 0.85 | | | | |
| 3 (crime rate) | 0.85 | | | | |
| 4 (diverse people) | | | 0.32 | | |
| 5 (where different people interact) | | | 0.44 | | 0.35 |
| 6 ("up and coming" vs. "established") | | | | | |
| 7 (get around by walking) | | 0.62 | | | |
| 8 (urban place) | | 0.33 | | | 0.47 |
| 9 (get around with public transit) | | 0.73 | | | |
| 10 (have a car) | | -0.67 | | | |
| 11 (looks nice) | 0.31 | | | 0.59 | |
| 12 (the look of the neighborhood) | | | | 0.93 | |
| 13 (bars and cafes) | | | | | 0.60 |
| 14 (speculate what it's like to live there) | | | 0.35 | | |
| 15 (do what locals do) | | | 0.62 | | |
| 16 (caters to tourists) | | | | | |
| 17 (off the beaten path) | | | 0.62 | | |
| 18 (functional neighborhood) | | | 0.33 | | |

Table 2. Factor loadings on survey questions, Airbnb data set. (findings on the general public data set were similar.) Loadings <0.3 are omitted. Factors 1 through 5 became safety, location convenience, living like locals, aesthetic appeal, and liveliness, respectively.

**Dimension 2: Location Convenience**
We define the "location convenience" of a place as how easy it is for a traveler staying at that place to get to everywhere they want to go. This concept emerged from our survey results; we had missed it in the interviews, in favor of the concept of "walkability." Walkability is part of location convenience: if a place is easy to walk around, then it will be easy to get to attractions and daily necessities. Especially in a foreign place, where one might not understand local public transport, walking is often the easiest way to get around.

However, location convenience extends beyond walkability, depending on the local transportation options and the traveler's trip. A1 and B1 both talked about the extensive subway in New York City; A1 noted that she did not feel compelled to stay in central Midtown as long as she was near a subway, while B1 preferred taking subways to walking because it was an interesting experience in itself. Perceptions of location convenience change with mode of transportation and circumstance, too. B5 and B6 described going to Aspen, Colorado together to see the Winter X Games extreme sports competition. Three areas, Aspen, Buttermilk, and Snowmass, had lodging options, but the only transport between them and the X Games site was by bus, and Buttermilk and Snowmass required an extra bus ride to get to Aspen. B5 and B6 focused their search on Aspen itself, because it would require one bus ride per day instead of two. Naturally, in a less crowded time of the year, Buttermilk and Snowmass would be more location convenient.

We did not ask specifically about location convenience in Study 2; its inclusion came out of the "other" responses. Of the 23 "other" responses from the Airbnb group, 13 were about location convenience, as were 2 of the 4 "other" responses in the general public group.

**Dimension 3: Living Like Locals**
This dimension represents how closely people can simulate a "normal", non-traveling life there, and approximates notions of "authenticity." Many participants expressed desires for an "authentic" "non-touristy" place. Clearly, "touristy" places have some disadvantages: they are expensive (B6 gave the example of paying £39 to see the Crown Jewels in London) and often people act differently there (B7 described feeling like she "had a dollar sign on her forehead" in the tourist beaches of Cancun). But those inconveniences do not explain the intensity of the desire to be "not a tourist" (or even "the anti-tourist", as A3 described himself). Furthermore, some people appreciated touristy places, for practical reasons: B7 noted that not speaking Spanish limited her experience in Mexico, and A2 described how she would search for a place that's not the #1 tourist destination but also not completely local, due to language issues.

To understand this tension, it is useful to review previous work about authenticity in tourist places. Early work located all spaces on a 6-stage scale from front-stage (purely for show) to backstage (fully authentic) and predicted that all tourists would seek authenticity [28]. Later work added more nuance, describing the "authenticity" of an experience in nine subtypes depending on how authentic the place was, how authentic the people were, and whether the visitor put importance on the authenticity of the people or the place, both, or neither [33]. Furthermore, the authenticity of an experience may be best explained as existential authenticity, or the personal resonance with that experience. Existential authenticity has two forms: intra-personal (discovering and being true to oneself) and inter-personal (having a real connection to others) [46].

We describe this nuance of authenticity because it reflects interviewees' actual usage. Many of them discussed "taking in the city life" (B1), seeing "what people actually do here" (A3), "kind of get[ting] a feel" of the city (A2), and even "play[ing] the game of, what if we lived here" (A7). When pressed, though, interviewees did not actually want their travel experiences to be about the real "everyday." Everyday life involves work, chores, and errands that most people do not enjoy, wherever they are. For example, asked if she would be interested to see everyday life in the Financial District of San Francisco, B1 replied no, the Financial District isn't the kind of "everyday" she's looking for (though clearly it is an integral part of many people's everyday lives). Instead, participants wanted to experience an "ideal everyday," which involved a relaxed day with plenty of third places.

Third places, such as bars, cafes, and bookstores as described in [32], are also a key part of this "ideal everyday." Many participants described local venues they loved: cafes where one can see friends sitting outside (A5), dive bars (B4). B1 went as far as to suggest that she would travel to a place based on where the best coffee shops were. Because third places tend to be neutral, accessible, status-leveling places, travelers appreciate them. Stepping into everyday life in another place involves adjustments, and these third places give travelers a way to recharge.

After our factor analysis of the survey results (see Table 2), we realized that the concepts of authenticity and third places, which we had assumed to be separate, really reflected the same underlying emotion: travelers want to be able to be part of a different place, to experience a different life instead of just seeing some different things. Therefore, we combined these two dimensions into one overall theme of "living like locals."

**Dimension 4: Aesthetic Appeal**
Aesthetic appeal in many forms is one of the main incentives for people to travel, and one of the main influences on the overall feeling of a trip. By "aesthetic appeal", we are referring to anything sensory: participants mentioned visual, auditory, and gustatory appeal particularly, and occasionally smell. Some preferences were universal, such as enjoyment of nature and avoidance of loud places while sleeping; others were more personal, like preferences for big cities or smaller towns.

**Dimension 5: Liveliness**
Our participants appreciated lively places. This intuitively makes sense, as they traveled to cities; the trips they discussed are not "sun and sand" getaway trips. Lively places have a number of advantages: there are usually businesses nearby, so it is easy to accomplish daily tasks; there are fewer safety concerns; they are often interesting in their own right due to markets or street performers. "Liveliness" encompasses what we previously referred to as "diversity" because the street is an equalizing place, as Jacobs writes [19]. All kinds of people can meet on a street; as A1 and B1 said, their favorite places allow "room for everyone."

Liveliness thus includes some of the components of diversity, some components of walkability, and some components of third places. Some participants offered examples of lively places they enjoyed: B1 enjoyed train stations, while she and B6 both brought up markets. Others described liveliness in their own words: "being in the middle of stuff" (B4) or being "where everything is" (A3). Liveliness makes traveling more pleasant and enables serendipitous encounters too. B6 talked about visiting New Orleans and stumbling across parades put together by local Native American groups, which she unexpectedly enjoyed. Similarly, B7 described how she preferred staying in the residential Itzimna neighborhood over the center of Merida, because she enjoyed her 20-30 minute walk to the center every day and the chance encounters it brings.

**Current neighborhood search tools are inadequate**
Currently, the primary search method participants said they used was to ask friends and family. If people visited friends, like B2 in Albuquerque and Portland, they can do this directly; otherwise, like A3, they would ask friends beforehand what were interesting and fun neighborhoods. Online research was also widely used, often as simply as Googling "things to do in London" or "London off the beaten path" (B6). B7 lamented, though, that this kind of searching can turn the normally-fun process of traveling into work.

Because searching was so labor intensive, some people who did not have any pre-existing heuristics (as described above) tried to create their own heuristics. A4 would search for the "queerest neighborhood" in a given city, as she did when she visited Zurich. This was not in order to find particular sites there, but just because she found that she would often like the kind of people she met there. Similarly, B1 searched for the best coffee shops, not because she would spend most of her time there, but because she usually likes neighborhoods that have good coffee shops. B2 would read books about a place, like Gregory David Roberts's novel *Shantaram* before visiting Mumbai, in order to recognize places they mentioned.

**DESIGN OF NEIGHBORHOOD GUIDES WEBSITE**
Based on this model of what creative tourists want, we built a prototype Neighborhood Guides web application as a probe for future studies. It is a website that travelers could visit while planning a trip, to gather information about neighborhoods they might stay in. A screenshot is shown in Figure 1. This site was not intended as a finished product, but rather as a probe to elicit reactions from participants to better understand what they need and want when traveling, and to see which side of the city is reflected by the social media that is posted there.

The information included is based on the five-dimensional model of creative tourist information search we built in the previous section. Each dimension is represented by a data source from social media or another publicly available source. Specifically, to address the dimension of safety, we use crime statistics from city open data portals; to address convenience, we include information from Walkscore[2]; to address aesthetic appeal, we include photos from Flickr; for liveliness, counts of venues of different kinds from Foursquare; for the ability to live like locals, common words from Twitter. (These are not all social media data sources, because social media can only satisfy some of users' needs. Again, we aim to understand how

---
[2]http://www.walkscore.com

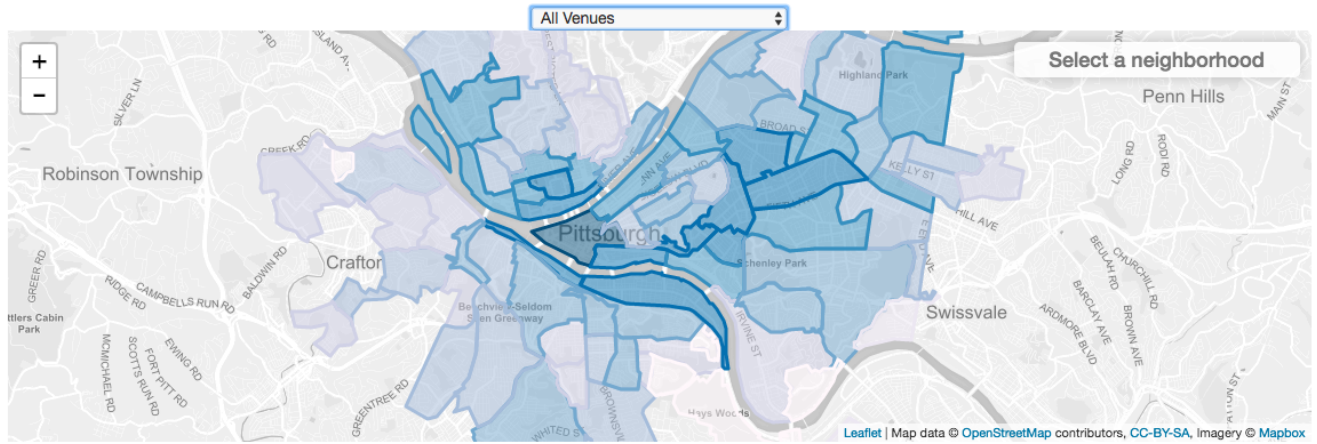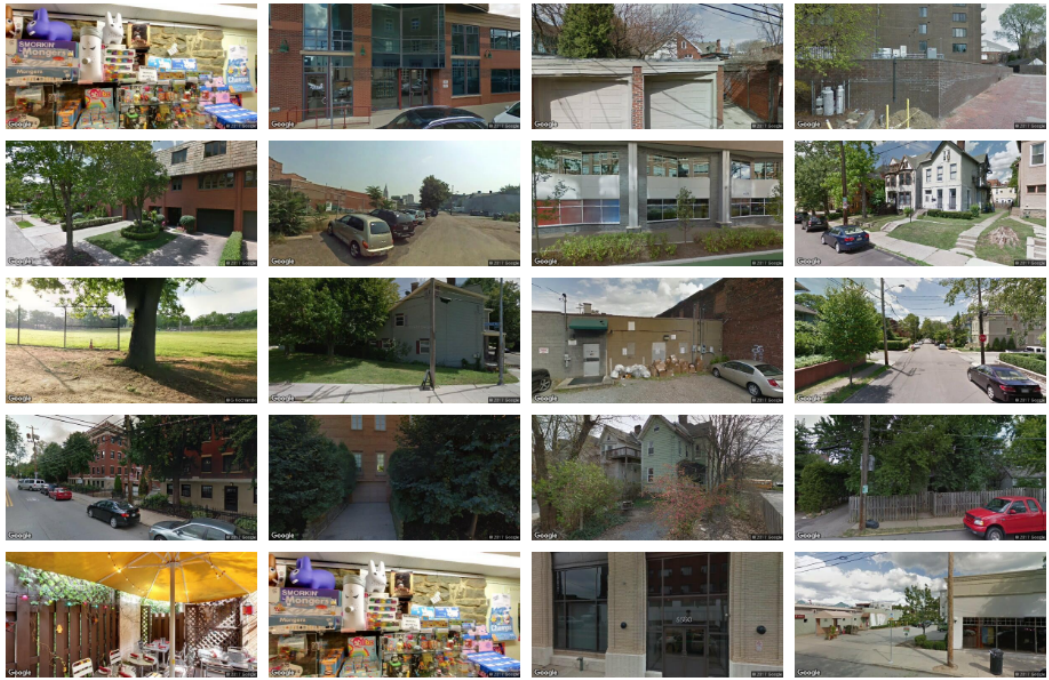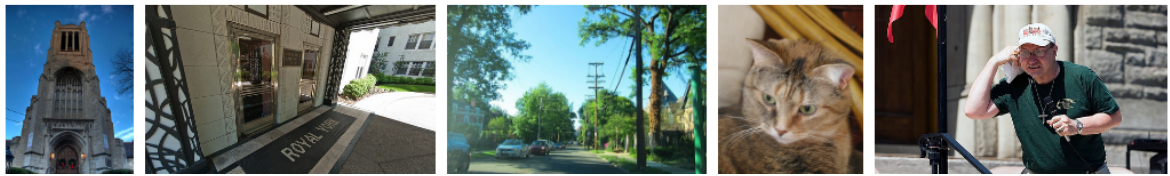

**Figure 1. A screenshot of our Neighborhood Guides prototype, version 2.0. The page continues below, with multiple sets of photos of the neighborhood from Flickr and Instagram. (We have only shared the first page here for space reasons.)**

| Section | Rank |
|---|---|
| Photos | 2.16 |
| Walk Scores | 2.33 |
| Venues | 3.00 |
| Crime | 3.67 |
| Tweets | 3.83 |

Table 3. Participants' average usefulness rank of each section in the initial evaluation. Lower is better (e.g. Rank=1 would mean that everyone ranked this section as the #1 most useful section).

social media can help these travelers, not to build a finished product using only social media.)

**City and Neighborhood Choices**
One important choice we had to make before beginning was which cities and neighborhoods to include. We chose Pittsburgh, San Francisco, Chicago and Houston because they were reasonably large cities from all sides of the country, and they had plenty of social media data available.

For each city, we used their open data portal to find neighborhood boundary data. The portals included the Western Pennsylvania Regional Data Center[3] for Pittsburgh, SF OpenData[4], City of Chicago Data Portal[5], and Houston Data Portal[6]. If they had multiple neighborhood boundary data sets, we selected the one that seemed to be the most canonical.

**STUDY 3: NEIGHBORHOOD GUIDES USER STUDY**
Our ultimate evaluation would help us find answers to our latter two research questions: can social media help creative tourists, and what does social media tell us about our cities? However, we first conducted an initial user study in order to catch any obvious mistakes and iteratively improve our prototype. This first evaluation was conducted with 9 travelers in public places in or near San Francisco. Participants were paid $15 for a session between 30-60 minutes in which we would talk with them briefly about their travel experience, describe the site, and ask them to use it to plan a hypothetical trip. After that, we would ask them, among other questions, to order the five parts of the site in usefulness. This study was approved by our university's Institutional Review Board.

**Photos are the most useful; Tweets are the least**
All users ranked the parts of the site. We scored rankings by giving 1 point to their most useful section, up to 5 points for their least useful section. If they found two equivalently useful, we gave them both the average of the two scores. (That is, if Photos and Venues were tied for first place, they would both get 1.5 points.) We then averaged across all 9 participants, ending up with the ranking in Table 3.

Photos were almost universally preferred as the most useful information source. Participants C5 and C6 described that the photos gave them the best flavor or feel; C7 described appreciating them because they were unfiltered. As she said, "it's easy to browse through and get a feel without having to read anything and get other people's opinions." Similarly, they have noticeable practical uses: in the hypothetical trip planning, C1 and C3 were both saved from neighborhoods that were more suburban and boring than they thought.

---
[3] http://www.wprdc.org
[4] https://data.sfgov.org
[5] https://data.cityofchicago.org
[6] http://data.ohouston.org

On the other hand, tweets were seen to be the least useful. C9 described how the words are useless without context, and sometimes the context provided failed to even give him the necessary context. Crime data was seen as almost as useless, but the distribution of its importance among participants is bimodal. As in Study 1, some described crime data as being the most important to them, while others did not care at all.

**STUDY 3, PART 2: USER STUDY, CONTINUED**
After the initial evaluation of the Neighborhood Guides, and the further study to show which photos are likely to be most helpful, we returned to the Neighborhood Guides application with more insights to guide our design. In this section, we describe first the improved Neighborhood Guides site, then detail the study we ran and the insights learned from it. We found that social media photos show the idealized side of the city that creative tourists want to see, that the best photos show people doing something, that statistics are useful but can be greatly simplified, and that people search for textual "blurbs" to give them a schema to base their understanding of the neighborhood on.

**Neighborhood Guides 2.0**
After the feedback from the initial evaluation, we made some changes to the Neighborhood Guides website. The biggest change was to focus on photos and hide tweets. This change was relatively straightforward, given our results in Study 3, part 1. Photos seemed the most immediately useful data source for travelers, and tweets were the least useful, so we reoriented our site to focus more on photos and we hid the tweets. The new site would only include the map and photos. In the user study we did with Neighborhood Guides 2.0, tweets were not shown. (We did retain the option to show them via a small control at the bottom of the page, for our use in the user study.) Of course, our results from the initial user study do not prove that there is no useful way to use geotagged tweets to describe an area. Our sample size was only 9 people, and even if a larger sample said the same thing, that would only prove that our method of selecting tweets did not describe an area well. Regardless, we chose primarily to use photos in order to focus our attention.

We showed the Street View photos first, as well as random Flickr photos with one per photographer. We maintained the original set of photos sorted by autotags, as well as Flickr photos selected according to the method in [20], and random Instagram photos, which started hidden but could be shown during the study.

**Study Methods**
We ran this user study as an extended version of Study 3, part 1. We would talk with the participants about a recent trip, then ask them to plan a hypothetical trip to one of the cities in our site that they had not yet been to. We asked 19 of the 21 participants (the other two ran out of time) about their preferences of photo subsets by showing each photo set.

We recruited participants through social media and via a participant pool at our university. Interviews lasted up to 60 minutes, and participants were paid $15. Sessions were conducted in a cafe at our university.

As in Study 1, state law and our IRB prevented us from recording these interviews. Instead, we took detailed notes and condensed them to the 2-7 most interesting findings after each session. We analyzed these notes using affinity diagramming (as described in [4]) to identify common themes.

### Results

*Flickr photos help travelers more than Street View*
When we asked users to rank which photos were the most useful to them, they responded as shown in Table 4. Flickr photos with categories were the best ranking, and Street View photos were the worst.

When asked why they ranked their choices the way they did, participants offered a few explanations. Most of them involved a variant of the idea that they don't actually want to see how a place really looks, they want to see something exciting about the place. D21 described wanting "to capture where people see beauty"; D1 described seeing "the best side of each neighborhood." Others described the Street View photos as boring (D7) or "too zoomed-in" (D6). D3, D4, and D12 appreciated the Flickr photos with autotag-based categories, as they help them process all these photos so quickly.

*The ideal photo has a person, doing a thing*
Many participants talked about wanting to learn not only what a place looks like, but also what people do there. If a photo just has a person's face, it is unhelpful; D7 reported wanting to see people doing things, "not just a picture of a guy." D21 also described how finding activities she could do in the place would be helpful. On the other hand, if a photo just has pictures of the place, that is likewise unhelpful; D3 and D12 described that being a shortcoming of the Street View photos. The ideal photo would show someone doing something; it would then represent something a traveler might do in the region.

On the other hand, photos of people participating in one-time events can be both positive and negative. D1 and D18 enjoyed learning about local events, but D8 and D10 noted that they are only helpful to travelers if the travelers happen to be in town when the event is happening. These events are often causes for taking lots of photos, too, which means that they may become the majority of the photos in a region. Therefore, when selecting photos from a neighborhood, half of the photos might come from something that happens only once a year. This was the case for D8, who found a lot of photos from the Washington Ave/Memorial Park neighborhood of Houston to be from a running event. She learned little about the neighborhood besides that it had some kind of annual race.

*A "blurb" gives people a conceptual start*
When researching a new neighborhood, participants struggled with trying to make sense of a lot of disparate information.

| Photo set | Average rank |
|---|---|
| FLICKR WITH CATEGORIES | 2.33 |
| FLICKR ONE-PER-USER | 2.57 |
| INSTAGRAM | 3.28 |
| FLICKR JAFFE | 3.30 |
| STREET VIEW VENUES | 3.53 |

Table 4. Users' (n=21) average rankings of photo sets in Study 3, Part 2. Lower is better; "1.00" would indicate that everyone ranked this set their #1 most useful set.

Seeing statistics on a map showed them one side of a neighborhood, but it was impossible to keep them all in their mind; similarly, the photos showed a lot of different stories from different people. They wanted to be able to tell one cohesive narrative about the neighborhood; whatever it was that all these photos and statistics had in common. Of course, any one narrative would naturally collapse a lot of the neighborhood's natural complexity, but that was fine; it would give them some way to organize all this information in their mind.

D10 described how the Neighborhood Guides site "does a lot at a glance; I kinda want to get the editorialized version." He described how he would read books, local blogs (like "I Heart Reykjavik[7]" when traveling to Iceland), books, or TV shows set in an area before traveling there. He used Wikitravel, an online travel guide, to see a factual overview of what's in a place; beyond that, he wanted some description.

Different travelers had different ways to find such a blurb. D15 used the Airbnb neighborhood guide's one-sentence overviews, like "Mexican bakeries, Chinese take out spots, artisanal donut shops, ramen restaurants, and lively bars all near Dolores Park.[8]" D2 and D20 both Googled the neighborhood they were looking at, settling on a local newspaper's page[9] and a tourism bureau site[10], respectively.

Lacking a short blurb, some participants would build their own picture based on a neighborhood's name. This is sometimes accurate, as when D13 was drawn to the Marina neighborhood in San Francisco because she likes being by the water, and when D6 rightly assumed Museum Park in Houston contained museums. However, it is often a mistake, as when D15 assumed "Russian Hill" was a heavily Russian area.

*Statistics are a rough search tool*
Participants used the map and statistics, but not thoroughly. They would use them to find where anything was, like D2 checking the "All Venues" or D6 trying to avoid "residential" places because there was not as much to do. They would also occasionally use the statistics to avoid high-crime areas, like D20 investigating Midtown, Houston. In these ways, they functioned as simple search tools, directing travelers not to one particular neighborhood but rather to a set of neighborhoods that were at least reasonably dense.

---

[7] http://www.iheartreykjavik.net/
[8] https://www.airbnb.com/locations/san-francisco/mission-district
[9] http://www.sfgate.com/neighborhoods/sf/hayesvalley/
[10] https://www.visithoustontexas.com/about-houston/neighborhoods/montrose/

## DISCUSSION

### Planning travel is getting excited and avoiding traps

Clawson and Knetsch, in 1966, described anticipation as the first part of an outdoor recreation experience. As they wrote, "a fisherman may get more enjoyment from tieing his own dry flies through the winter than he will later get from the actual fishing itself" [6]. This principle rings true in our participants' stories and plans, and can explain many of our findings.

A photo showing someone doing an activity would inspire much more anticipation than one simply showing a building or a person; the benefit of the photos is that they can mentally bring people into that activity. Picture quality and page polish matters (D4 skipped over some sections because they had missing photos; D9 selected as her most useful photo set the one that had the aesthetically best photos). If participants were trying to understand some "true" perception of the city, the photo quality would not matter. But if they are trying to get excited about the area, of course they want to see the most beautiful side of each neighborhood. (Incidentally, this factor may also help explain why Street View photos did better in the Mechanical Turk study. They are all the same size, so we could display them in a neat grid, while the social media photos could not be so neatly arranged.)

In addition, the short "blurb" that the participants sought out makes more sense if their goal was to find an exciting part of a city than if they were trying to find an accurate view. No short blurb can explain, for example, all of the history, culture, problems, and triumphs of San Francisco's Mission District, but a blurb can quickly point readers to its "Mexican bakeries, Chinese take out spots, ... and lively bars near Dolores Park."

There is another side to the trip-planning process, however: the necessity of avoiding traps. There are a number of kinds of traps, including unusually high-crime areas, surprisingly spread-out residential neighborhoods, or quintessential tourist traps. This is why D13 was interested in "authenticity;" despite not being able to define it exactly, she knew that "staged" photos from tourist attractions might lead her into a trap. This is also why D19 and D20 changed their plans from their first inclinations (Tenderloin in San Francisco and Midtown Houston); the areas looked interesting, but the participants realized they had a bit more crime than they expected.

Thus, travelers must manage this tension of building excitement while avoiding traps. First person accounts and photos build excitement; by reading individual stories, photos, and blog posts, a traveler can see all the compelling scenarios that might play out when they travel there. However, it is much easier to find what not to do by taking a wider view like that offered by statistics. In the rest of this section, we will explain how these two components can do their jobs optimally.

### Social media photos can help travelers get excited

The role social media photos can play for travelers involves building their anticipation about a place. This is why they were the best choices when we asked "what photos are most useful when you're traveling?", while the Street View photos were ranked higher in the Mechanical Turk study when we asked "which photos best represent your neighborhood?" Our participants liked seeing people doing things, they liked seeing a diverse array of photos, and they liked seeing the most beautiful and well-shot photos in each neighborhood.

The question of which photos should be shown is still a promising open research question. We did not find evidence to support the algorithm of [20], but we did not find any other photo set that was consistently more effective either. We hope further research can develop a consistently successful algorithm. In the meantime, we did uncover some suggestions for current application developers. Low-quality, low-resolution, or badly-lit photos should be removed. Photos from the same photographer or day should be limited; photos are more useful to travelers when they represent a diverse cross-section of the experiences in the area. Some kind of organization, like our autotag-based categorization, seems useful.

### Statistics can help travelers avoid traps

Statistics still play an important role in helping travelers, but it is a niche role. Our statistics could be further simplified because they are mostly just proxies for residential density: higher downtown, lower in outlying neighborhoods. Clearly the venue densities are this way; Walk Scores tend to follow this pattern too, because it's easier to walk, bike, or take transit when everything is closer together. Crime per capita is not always higher downtown, but it often is; D5 explicitly mentioned that when he saw high crime in the Loop of Chicago.

Therefore, we could replace all of the chloropleth maps with a simple map that showed residential density. This would answer the most common question people used the map for: "where is everything?" This would easily help them avoid the pitfall of booking a place that happens to be in an inaccessible or boring neighborhood, while not overwhelming them with numbers. Perhaps a crime map could be included too; as we found in our first exploratory study, it's something that most people don't care about, but a few people cannot live without.

As another option, we could look to the work of Correll and Heer [7]. They describe a Bayesian algorithm used to generate "Surprise Maps" that show how surprising a statistic in an area is. This would fit perfectly with our needs: instead of showing crime rates, for example, the map could show only areas that have surprisingly high crime, helping visitors avoid that area. They could also provide a window into interesting areas: knowing an area has a lot of shops is not interesting, but knowing it has more than its fair share of record stores or auto repair shops can tell a traveler a lot about the place.

## CONCLUSION

A new generation of tourists has arrived, but the tools they need have not. Geotagged social media can help fill this gap and help them understand the neighborhoods in an unfamiliar city, and we have provided an example and guidelines to build a better site. In doing so, we have also added to our knowledge about a bigger question: "What does social media tell us about our cities?" Social media shows us the idealized view of the city, each person's own highlights of their life there. We hope that these insights will both help tourism developers make better tools, and help social media researchers better understand the available data.